\documentclass[12pt,letterpaper]{article}
\usepackage{amsmath}
\usepackage{amssymb}
\usepackage{cite}
\usepackage{array}
\usepackage{setspace}
\usepackage{graphicx}
\usepackage{endfloat}
\usepackage{booktabs}
\def\d{{\mathrm{d}}}

\def\br{{\bf r }}

\def\adf{{\sc ADF}}
\def\adftext{{\sc Amsterdam Density Functional}}

\def\GG{{(GG)$^+$}}
\def\Ad{{(AA)$^+$}}
\def\AG{{(AG)$^+$}}
\def\TG{{(TG)$^+$}}
\def\GA{{(GA)$^+$}}
\def\GT{{(GT)$^+$}}
\def\TT{{(TT)$^+$}}
\long\def\symbolfootnote[#1]#2{\begingroup%
\def\thefootnote{\fnsymbol{footnote}}\footnote[#1]{#2}\endgroup} 

\begin{document}

\begin{center}
{\large\bf An Accurate and Linear-Scaling Method for Calculating Charge-Transfer Excitation Energies and Diabatic Couplings
}\\[2ex]
{\large Michele Pavanello\footnote{E-mail: m.pavanello@rutgers.edu}}\\
Department of Chemistry, Rutgers University, Newark, NJ 07102-1811, USA \\[10pt]
{\large Troy Van Voorhis}\\
Department of Chemistry, Massachusetts Institute of Technology, \\ 
Cambridge, MA 02139-4307, USA\\[10pt]
{\large Lucas Visscher}\\
Amsterdam Center for Multiscale Modeling, VU University, \\
De Boelelaan 1083, 1081 HV Amsterdam, The Netherlands \\[10pt]
{\large and} \\[10pt]
{\large Johannes Neugebauer\footnote{E-mail: j.neugebauer@uni-muenster.de}}\\
Theoretische Organische Chemie, 
Organisch-Chemisches Institut der
Westf{\"a}lischen Wilhelms-Universit{\"a}t M{\"u}nster\\
Corrensstra{\ss}e 40, 48149 M{\"u}nster, Germany\\[2ex]
\vfill
\end{center}
\begin{tabbing}
Date:   \quad\= \today \\
Status: \> {\it J. Chem. Phys.} accepted for publication\\
\end{tabbing}
\newpage
\begin{abstract}
Quantum--Mechanical methods that are both computationally fast and accurate are not yet available for electronic excitations having charge transfer character. 
In this work, we present a significant step forward towards this goal for those charge transfer excitations that take place between non-covalently bound molecules. In particular, we present a method that scales linearly with the number of non-covalently bound molecules in the system and is based on a two-pronged approach: The molecular electronic structure of broken-symmetry charge-localized states is obtained with the Frozen Density Embedding formulation of subsystem Density-Functional Theory; subsequently, in a post-SCF calculation, the full-electron Hamiltonian and overlap matrix elements among the charge-localized states are evaluated with an algorithm which takes full advantage of the subsystem DFT density partitioning technique.
The method is benchmarked against Coupled-Cluster calculations and achieves chemical accuracy for the systems considered for intermolecular separations ranging from hydrogen-bond distances to tens of {\AA}ngstroms. Numerical examples are provided for molecular clusters comprised of up to 56 non-covalently bound molecules. 
\end{abstract}
\newpage
\section{Introduction}
Charge transfer (CT) reactions are ubiquitous in all branches of chemistry, and huge efforts have been spent in the past to analyze these processes in terms of theoretical models \cite{MayKuehn_book,KuznetsovUlstrup_book,Nitzan_book}.
They can be categorized into process in which a charge separation in a donor--acceptor system takes place,
\begin{eqnarray}
\label{pm}
D + A &\to& D^+ + A^-,
\end{eqnarray}
and processes where the charge has been externally created (either by injecting or removing an electron),
\begin{eqnarray}
\label{m}
D^- + A &\to& D + A^-, \\
\label{p}
D^+ + A &\to& D + A^+,
\end{eqnarray}
where we define $D$ as the donor and $A$ as the acceptor. For the reactions in (\ref{pm}) and (\ref{m}), donor and acceptor are defined in terms of donating and receiving electrons, while in (\ref{p}) the definitions are in terms of the location of the excess positive charge.
The three kinds of CT reactions in (\ref{pm}--\ref{p}) are involved in a plethora of important processes. For example (\ref{pm}) may resemble a charge splitting event either at a semiconductor interface, between an organic dye and a semiconductor, or the charge separation event in the reaction centers of photosynthetic systems. Generally, all the charge separation events can be thought of in terms of a two-state model comprised of a neutral state and a charge separated state.
Throughout this work, we call the reactions involving only the motion of an excess of charge, as in (\ref{m}) and (\ref{p}), migration CT (MCT, hereafter) reactions. MCT reactions are the simplest type of reactions involving CT states, and take place in many fundamental processes related to charge mobility, transport, and enzyme functionality. In these processes, the charge separation event has happened in the past, and the problem shifts to the prediction of the kinetics of the excess charge, may that be across a polymer \cite{mill2004,inze_book,reut2012} (organic electronics), a protein  (enzyme functionality \cite{grey1996,kawa2006} and photosynthesis \cite{rome2009a,scho2011}), or DNA \cite{stor2001,gies2000,gene2010} (oxidative damage).

Very often, the ground and MCT states are close in energy and one easily finds the first MCT state as being the first excited state of the system. This is an important advantage in theoretical studies of this type of states as opposed to other types of excited states for which the excited state search can be a daunting task. Another important quality of MCT states is the simple physical depiction of ground and excited states: the former has an excess charge on the, say, ``left'', and the latter features an excess charge on the ``right''. Obviously, this simple depiction becomes more complicated as soon as the system under study contains more than two chemical moieties capable of accepting/donating an excess charge. These unique qualities make MCT states a perfect testbed for new computational methods aiming at the accurate prediction of generally all kinds of CT excitations.

There are two fundamentally different schemes for calculating CT excitations. The first one involves the construction of Hamiltonian and overlap matrices in a basis of charge-localized broken-symmetry states. This basis set is often called a diabatic basis set, and the basis functions termed diabatic states. The chosen broken-symmetry states in the set should resemble the true ground and the CT excited states of the system. Depending on the size of the basis set (or equivalently the number of broken-symmetry states included in the calculation) the simultaneous diagonalization of overlap and Hamiltonian matrices (i.e.\ solving the generalized eigenvalue problem) yields the sought ground and CT excited states. The latter diagonalization step is similar to the diagonalization step in the Configuration Interaction (CI) method carried out with valence bond configuration state functions. Even though this is a general method, it is particularly suited for the calculation of CT excited states. This is because contrary to valence excitations, the broken-symmetry basis functions assume a simple charge-localized character and relatively uncomplicated computational criteria can be devised for their generation \cite{kadu2012,vanv2010a,hsu2009,hong2006,cave1997,subo2010,migliore2009b}. This is the method that we use in this work and is described in detail in the next section for a basis set comprised of two broken-symmetry states.

Contrary to the first scheme mentioned above, the alternative is not tailored specifically to CT excitations and it involves obtaining directly the adiabatic ground and excited states of the system. As such it is more general than the first scheme and is most often adopted in the literature. Due to the high computational throughput of modern computers, calculations based on the latter scheme are now commonplace. A plethora of quantum chemistry methods are devoted to the prediction of excited states' energies, transition moments, densities, and wave functions. Despite this, CT excited states have always been more challenging than others, because for this kind of excitations the electron density changes dramatically compared to the ground state one. Density-functional theory (DFT), and specifically its time-dependent extension (TD-DFT) through the linear-response formalism for the calculation of electronic excitations and transition moments~\cite{casi1995} has struggled to correctly deal with CT excitations when employing approximate density functionals~\cite{dreu2005}. Pragmatic corrections exist, such as the one originally developed by Gritzenko {\it et al.}~\cite{grit2004,neug2006a}, or the one involving the {\it Peach factor}~\cite{plotn2010}. Range separated~\cite{yana2004,toul2004,gerb2005} density functionals have also offered an effective solution to the problem by enforcing the correct long-range behavior at the price of including yet another parameter to the density functional (the range separation parameter, often denoted by $\gamma$). Methodologies to obtain $\gamma$ parameters which satisfy certain system-dependent and process-dependent conditions have been proposed~\cite{livs2008,baer2010}.
Novel computational approaches to obtain charge-transfer excited states within a DFT formalism are being explored. Specifically, a variational formulation of TD-DFT~\cite{zieg2010} has shown to alleviate or to completely cure the CT excitation failures of linear-response TD-DFT when the fourth-order relaxed constrained-variational TD-DFT method is employed, however, at the expense of higher computational complexity than standard linear-response TD-DFT.

Wave function based methods also have encountered difficulties when approaching CT excitations. The low end of wave function methods, configuration interaction with singles (CIS) and time-dependent Hartree--Fock (TD-HF), are known to grossly overestimate CT excitation energies~\cite{subo2011,dreu2003,dreu2004,auts2009,grim2007}. More balanced methods, such as multi-configuration self-consistent field (MCSCF), multi-reference CI (MR-CI), or perturbative corrections to CIS, such as CIS(D), as well as ADC(2)~\cite{hatt2005} are able to deliver good CT excitation energies~\cite{subo2011,duto2011} and are often taken as benchmark~\cite{aqui2011}. MCSCF, however, is known to fail near avoided crossings (a feature always present in systems featuring MCT excitations) unless state averaging is employed~\cite{kryl2006}. The equation-of-motion Coupled Cluster (EOM-CC) theory~\cite{stan1993,kryl2006} is also known to fail in reproducing CT excitations with a similar accuracy than ionization potentials and electron affinities~\cite{stan1994,musi2011} due to its non size-extensivity stemming from the non-exponential from of the EOM-CC ansatz for the excited states. Linear-response CC, and specifically an approximation to it including only single and double excitations known as CC2 has been very successful in predicting CT excitations~\cite{wong2008,stei2009a,stei2009b,hatt2006}. A particularly powerful implementation of CC2 utilizing the resolution of the identity (RI-CC2) \cite{hatt2000} is routinely applied to systems with up to 100 atoms. It is now understood that to obtain an accurate description of CT excitations the employed method must describe the dynamic correlation of the CT excited states similarly to the one of the ground state~\cite{glae2010}. In addition, the computational costs associated with all the wave function methods mentioned above (with the exception of TD-HF and CIS) are prohibitive for most systems in condensed-phase and biosystems of interest.

In the next section, we will describe in detail how ground and CT excited state energies and wave functions can be obtained starting from a basis set comprised of two broken-symmetry states. In Section \ref{sect_FDE} we will introduce the Frozen Density Embedding fomulation of subsystem DFT (a linear-scaling, full-electron electronic structure method) which we use to determine the electronic structure of the broken-symmetry, charge-localized states. Then in section \ref{sect_CTEFDE} we will present the theory behind the calculation of the full-electron Hamiltonian and overlap matrix elements among the broken-symmetry states (generated with subsystem DFT) which we have implemented in the \adftext\ (\adf ) computer software~\cite{teVelde:2001a,jaco2008b}. Section \ref{sect_Val} is devoted to benchmarking of the method for two selected cases of MCT, e.g.\ hole transfer in water and ethylene dimers at several inter-monomer separations. Section \ref{sect_pilot} features two pilot calculations of the hole transfer in water and ethylene clusters containing up to 56 and 20 molecules, respectively.

\section{Excitation energies from a model of two broken-symmetry states}

Consider a system comprised of a charge donor and a charge acceptor moiety in the absence of low-lying intermediate bridge states (DA system hereafter). Such a system can be well characterized by a two-state formalism. That is, it is enough to consider either the adiabatic ground and first excited state or a set of two broken-symmetry charge-localized states to capture most of the underlying physics. There is a large literature \cite{newton1991,migliore2011a,hsu2009,plass2011,vanv2010a} 
supporting the idea that the states with localized excess charges, i.e.\ one where the charge is localized on the donor and, conversely, one where the charge is localized on the acceptor, are an appropriate basis for modeling the process of charge migration between donor and acceptor.  

A state with localized charges generally is not the ground state of the system Hamiltonian. Therefore, in the basis of charge-localized states, the Hamiltonian matrix is non-diagonal. In addition, as these states are of broken-symmetry character, they are generally non-orthogonal to each other.
Throughout this work, we will refer to the full-electron wave functions of the two charge-localized states as $\Phi_1$ and $\Phi_2$. 
For DA systems, the Hamiltonian ($\mathbf H$) and overlap matrices ($\mathbf S$) in this basis are of dimension $2\times 2$, namely
\begin{equation}\label{Hdia}
\mathbf{H}=\left(
\begin{array}{cc}
 H_{11} & H_{12}\\
 H_{21} & H_{22} 
\end{array}
\right),~\mathrm{and}
\end{equation}
\begin{equation}\label{Sdia}
\mathbf S=\left(
\begin{array}{cc}
 1      & S_{12}\\
 S_{12} & 1 
\end{array}
\right).
\end{equation}
The definitions in Eqs.\ (\ref{Hdia}) and (\ref{Sdia}) allow us to obtain the CT excitation energy as the energy difference of the orthonormal adiabatc states, which can be obtained by solving the generalized eigenvalue problem, 
\begin{equation}\label{GEPdia}
\left| 
\begin{array}{cc}
 H_{11}-E        & H_{12}-E S_{12}\\
 H_{21}-E S_{12} & H_{22}-E 
\end{array}
\right|=0.
\end{equation}
The above equation yields two energy eigenvalues (say $E_1$ and $E_2$) and their difference ($\Delta E$) is
the sought CT excitation. In a closed form, 
we obtain \cite{migliore2011a,efri1976,newton1991}
\begin{equation}\label{Eex}
\Delta E = \sqrt{ \frac{\left(H_{11}-H_{22}\right)^2}{1-S_{12}^2} + 4 V_{12}^2 },
\end{equation}
where $V_{12}=\frac{1}{1-S_{12}^2}\left[ H_{12}-S_{12}\frac{H_{11}+H_{22}}{2} \right]$.

The problem is then shifted to calculating four matrix elements, i.e.\ the two diagonal Hamiltonian elements ($H_{11}$ and $H_{22}$, e.g.\ the total energies of the charge-localized states or diabatic energies), the $H_{12}$ off-diagonal matrix element, and the $S_{12}$ overlap element. The $V_{12}$ element introduced above is commonly referred to as electronic coupling.

Several methods and algorithms have been proposed to generate ad-hoc charge-localized states and to calculate the above matrix elements both for purely wave function methods
\cite{hsu2009,subo2010,cave1997,mo1998,gian1996,voityuk2002} as well as DFT methods
\cite{kadu2012,vanv2010a,wu2006,migliore2011a,kubar2008a,hong2006,voityuk2001,pava2011b}. 
Regular wave function and DFT methods tend to scale non-linearly with the system size. Therefore, important systems of biological interest, large organics and hybrid organic--inorganic systems (such as dye-sensitized solar cells) are largely out of reach of these methods unless truncated model systems or substantial approximations are introduced at the electronic structure theory level.

In this work, we build upon ideas presented in a previous paper~\cite{pava2011b} and we construct the broken-symmetry charge-localized states using a linear-scaling technique called Frozen Density Embedding~\cite{weso1993}. The construction of the charge-localized, diabatic states is a prerequisite in this formalism for the calculation of the CT excitation energies. A brief description of this technique follows this section. 
%
%
%
%
\section{The Frozen Density Embedding Formulation of Subsystem DFT}
\label{sect_FDE}
Subsystem DFT is a successful alternative to regular
Kohn--Sham (KS) DFT methods due to its ability to overcome the computational difficulties 
arising when tackling large molecular systems\cite{cort1991}. 
One particular variant of subsystem DFT is the FDE approach developed by Wesolowski and Warshel~\cite{weso1993}.
FDE splits a system into interacting subsystems and yields subsystem electron densities separately. Hence, it defines the total electron density, 
$\rho(\br)$, as a sum of subsystem densities~\cite{weso1993},
\begin{equation}
\label{eq:int:1}
\rho(\br)=\sum_I^{\textrm{\# of subsystems}} \rho_I (\br),
\end{equation}
where the sum runs over all subsystems. 
The subsystem densities are found by solving subsystem-specific 
KS equations, known as KS equations with constrained electron 
density or KSCED \cite{weso2006}. In these equations, the KS potential $v_{KS}(\br)$ is
augmented by an embedding potential $v_{emb}(\br)$ which includes
the electrostatic interactions taking place between electrons and nuclei 
of the subsystems as well as a potential term deriving from the non-additive
kinetic energy ($T_s^{\rm nadd}$) and non-additive exchange-correlation 
($E_{xc}^{\rm nadd}$) functional derivatives. 
We refer to Refs.\ \cite{weso1993,weso2006,solo2012}
for more details on the FDE theoretical framework, however, for sake of clarity we report here the spin KSCED equations which lead to the subsystem orbitals
\begin{equation}
\label{KSCED}
\left[ \frac{-\nabla^2}{2} + v_{\rm KS}^{I\sigma}(\br) + v^{I\sigma}_{emb}(\br) 
\right]\phi_{(i)_I}^{\sigma}(\br) = \epsilon_{(i)_I}^{\sigma} \phi_{(i)_I}^{\sigma}(\br),
\end{equation}
where $\phi_{(i)_I}^{\sigma}$ are the molecular orbitals of subsystem $I$ and of spin $\sigma$.

The FDE scheme greatly reduces the computational cost compared to KS-DFT, 
as there is no need to calculate orbitals for the total (``super'') system,
and the total computational complexity is then linear over the number of subsystems 
composing the total supersystem. 
%
%
{ However, the reduced cost in FDE compared to KS-DFT is achieved at the expense of 
approximating the non-additive functionals with semilocal density-functional approximants. This approximation, for the kinetic 
energy especially, is the single source of certain shortcomings of FDE, for example when applied to covalently bound subsystems \cite{fux2010}.}
%
%
%
\section{Charge Transfer Excitations with FDE}
\label{sect_CTEFDE}
Recently, two of the present authors have shown in Ref.\ \cite{pava2011b} that FDE can be successfully used to calculate migration CT couplings and excitation energies. In this work, we present a different algorithm for obtaining the elements of the full-electron Hamiltonian and overlap matrices defined in Eqs.\ (\ref{Hdia}) and (\ref{Sdia}) which has significant advantages over the previous method. Specifically, the new formalism can be applied to symmetric systems (out of reach before due to a singularity in the working equations \cite{pava2011b,migliore2009b}) and, as it will be clear in Section \ref{sect_Val}, yields very accurate excitation energies with mainstream GGA-type functionals, while before it was necessary to include HF non-local exchange to the functional to counteract the self-interaction error. In addition, two new algorithms have been implemented here. We call them subsystem-joint transition density (JTD) and subsystem-disjoint transition density (DTD) formalisms. While the former is computationally straightforward, it formally does not scale linearly with the number of subsystems. Instead, the latter scales linearly with the number of subsystems and, thus, it is theoretically consistent with the FDE formalism.

\subsection{Approximate Couplings and Total Energies in Subsystem DFT}
\label{sect_coupl}
Suppose we have two broken-symmetry Slater determinants describing two charge-localized states 
$\Phi_1$ and $\Phi_2$. The wave functions in terms of the two set
of molecular orbitals take the form
\begin{equation}\label{phidef}
\Phi_i=\hat A \left[ \phi_1^{(i)} \phi_2^{(i)} \ldots \phi_N^{(i)} \right],
\end{equation}
where the antisymmetrizer $\hat A$ also contains appropriate normalization constants.
Generally, the two sets of molecular orbitals $\{\phi_k^{(1)}\}$ and $\{\phi_k^{(2)}\}$
may not be orthonormal to each other and within the sets. This usually happens for broken-symmetry HF and KS-DFT 
solutions as well as for KS-determinants derived from 
Constrained DFT~\cite{kadu2012} and subsystem DFT calculations \cite{pava2011b}. We define the transition orbital overlap matrix as follows
\begin{equation}\label{overdef}
\mathbf (\mathbf S^{(12)})_{kl}=\langle \phi^{(1)}_k | \phi^{(2)}_l \rangle.
\end{equation}
In what follows, we only consider the case of a subsystem DFT 
calculation, i.e.\ we only deal with Slater determinants 
of the system constructed form subsystem molecular orbitals as 
done in a previous work \cite{pava2011b}. 
To avoid redundance with the theory of Ref.\ \cite{pava2011b}, let us briefly state that, similarly to the Valence Bond theory \cite{gian1996}, the subsystem DFT version of the Slater determinant in Eq.\ (\ref{phidef}) features products of occupied subsystem orbitals regardless of the fact that the orbitals between subsystems might not be orthogonal to each other.
In the case of a two-subsystem partitioning, the $\mathbf S^{(12)}$ transition orbital overlap matrix can be formally written as
\begin{equation}\label{overdef2}
\mathbf S^{(12)} =
\left( 
\begin{array}{cc}
\mathbf S^{(12)}_I & \mathbf S^{(12)}_{I,II} \\
\mathbf S^{(12)}_{II,I} & \mathbf S^{(12)}_{II} 
\end{array}
\right),
\end{equation}
where $\mathbf S^{(12)}_{I}$ and $\mathbf S^{(12)}_{II}$ are transition orbital overlap matrices calculated with the 
orbitals belonging to subsystem $I$ or $II$, respectively (subsystem transition orbital overlap matrices, hereafter),
while $\mathbf S^{(12)}_{I,II}$ and $\mathbf S^{(12)}_{II,I}$
include the overlaps of the orbitals belonging to subsystem $I$ with the ones
belonging to subsystem $II$. 
Let us clarify that there are two sources of non-orthogonality in this formalism. The first one stems from the overlap between the full-electron charge-localized states, i.e.\ Eq.\ (\ref{Sdia}), and the second one arises at the (subsystem) molecular orbital level in Eqs.\ (\ref{phidef}) and (\ref{overdef}). The orbital overlap matrix is generally non-diagonal, also
 when it is computed from orbitals belonging to a single state, namely, 
\begin{equation}\label{overdefdiag}
(\mathbf S^{(ii)})_{kl}=\langle \phi^{(i)}_k | \phi^{(i)}_l \rangle.
\end{equation}
It can be seen that the above matrix also is non-diagonal as the orbitals $\{ \phi^{(i)}_k \}_{k=1,N}$ are borrowed from an FDE calculation which does not impose orthogonality to orbitals belonging to different subsystems \cite{weso1993,iann2006}.

Going back to the elements of the $\mathbf S^{(12)}_{I,II}$ and $\mathbf S^{(12)}_{II,I}$ submatrices, it is clear that they
are small in magnitude compared to the subsystem transition overlap matrices provided that the subsystems are not 
covalently bound to each other.
The matrix form in Eq.\ (\ref{overdef2}) partially loses its block form in the case
of electronic transitions that change the number of electrons in the subsystems, such as the ones considered in this work. Specifically, if the CT transition involves one electron, then the transition overlap matrix will have one column (row) with sizable non-zero elements across the subsystems involved in the CT event.

In this work, we use the following formulas for the calculation of the Hamiltonian coupling between the two charge-localized states~\cite{kadu2012, thom2009a}
\begin{equation}\label{newcoupling}
H_{12}=\langle \Phi_1 | \hat H | \Phi_2 \rangle = E\left[ \rho^{(12)}(\br) \right] S_{12}
\end{equation}
where $S_{12}=\det\left( \mathbf S^{(12)} \right)$, and 
\begin{equation}\label{supratrans}
\rho^{(12)}(\br)=\sum_{kl} \phi^{(1)}_k (\br)\left(\mathbf S^{(12)} \right)^{-1}_{kl}  \phi^{(2)}_l (\br)
\end{equation} 
is a scaled transition density derived from the orbitals of the two states, and the functional $E$ is an appropriate density functional. The transition density in Eq.\ (\ref{supratrans}) is obtained from the integration of the one-particle Dirac delta function, namely~\cite{maye2003ch5}
\begin{align}
\nonumber
\langle \Phi_1 | N \delta (\br_1 - \br) | \Phi_2 \rangle &=N \int \d \br_1 \cdots \d \br_N \Phi_1(\br_1 \cdots \br_N) \delta (\br_1 - \br) \Phi_2(\br_1 \cdots \br_N)\\
\nonumber
&=\sum_{kl} D^{(12)}(k|l) \phi^{(1)}_k (\br)  \phi^{(2)}_l (\br)\\
\label{supratrans2}
&=\det\left( \mathbf S^{(12)} \right) \sum_{kl} \phi^{(1)}_k (\br)\left(\mathbf{S}^{(12)} \right)^{-1}_{kl}  \phi^{(2)}_l (\br)
\end{align}
where $D^{(12)}(k|l)$ is the signed minor (or cofactor) of the transition orbital overlap matrix of Eq.\ (\ref{overdef}).
Then Eq.\ (\ref{supratrans}) is obtained by simply dividing Eq.\ (\ref{supratrans2}) by 
$\det\left( \mathbf S^{(12)} \right)$, so that the scaled transition density integrates to the total number of electrons ($N$) rather than to $N \det\left( \mathbf S^{(12)} \right)$. 
%

%
%
{Scaling the transition density is a matter of algebraic convenience. Eq.\ (\ref{newcoupling}) is particularly simple in terms of a the scaled transition density, and is rigorous if the wavefunctions are single Slater determinants and the Hamiltonian is the molecular electronic Hamiltonian \cite{fara1990,king1967,mcweeny,thom2009a} (i.e.\ as in the HF method).}
%
%
%
%
{In our work, we consider two charge-localized states per calculation. These are generally non-orthogonal. However, for some systems, it is possible that the two states could accidentally be orthogonal. In this scenario, the scaled transition density is recovered by employing the Penrose inverse of the transition overlap matrix in Eq.\ (\ref{supratrans}). The matrix is inverted in the $N-x$ dimensional subspace (usually $x=1$) where is it is not singular.}
%
%

It is important to point out that  the coupling formula in Eq.\ (\ref{newcoupling}) was derived assuming that $\Phi_{1/2}$ are broken-symmetry HF wave functions. Even though the HF and KS wave functions are both single Slater determinants, the formula is not formally transferable from the HF method to a DFT method (as we do in this work). Applying Eq.\ (\ref{newcoupling}) in the context of DFT is an approximation. It can be shown that in the context of linear-response TD-DFT  Eqs.\ (\ref{newcoupling}) and (\ref{supratrans}) are equivalent to the Tamm-Dancoff approximation to TD-DFT.
In addition, let us clarify that formally the HF exchange and HF and KS kinetic energy terms are computed with the transition density {\em matrix}, and thus the density in Eq.\ (\ref{supratrans}) should be replaced by a quantity that depends on one electronic coordinate $\br$ as well as another electronic coordinate, $\br^{\prime}$, namely $\rho^{(12)}(\br,\br^{\prime})=\sum_{kl} \phi^{(1)}_k (\br)\left(\mathbf S^{(12)} \right)^{-1}_{kl}  \phi^{(2)}_l (\br^{\prime})$. However, for sake of clarity, here we will limit ourselves to reporting the (transition) density without introducing the density matrix notation.
As an example of how we apply Eq.\ (\ref{newcoupling}) in practical calulations, consider a calculation in which we employ the LDA exchange density functional, then the exchange contribution to the off-diagonal Hamiltonian matrix element becomes
\begin{equation}
E_{\rm x}\left[ \rho^{(12)}(\br) \right] S_{12}= - S_{12} \frac{3}{4}\left( \frac{3}{\pi} \right)^{\frac{1}{3}}
\int \left( \rho^{(12)}(\br) \right)^{\frac{4}{3}} \d\br.
\end{equation}

The diagonal elements of the Hamiltonian are computed in a similar fashion as the off-diagonal ones, namely
\begin{equation}\label{newener}
H_{ii}=\langle \Phi_i | \hat H | \Phi_i \rangle = E\left[ \rho^{(i)}(\br) \right] 
\end{equation}
where
\begin{equation}\label{supradens}
\rho^{(i)}(\br)=\sum_{kl} \phi^{(i)}_k (\br)\left(\mathbf S^{(ii)} \right)^{-1}_{kl}  \phi^{(i)}_l (\br).
\end{equation} 
Note that the overlap element has disappeared in Eq.\ (\ref{newener}). While this is a trivial consequence of the normalization condition on the Slater determinant in a regular KS-DFT calculation, the KS Slater determinant of the supersystem built with subsystem orbitals is not normalized. However, the corresponding FDE density theoretically is the correctly normalized correlated density of the full system, integrating to the total number of electrons~\cite{pava2011b}. This is an important point as the number of electrons in a calculation is set by the trace of the (transition or diagonal) orbital overlap matrix, while the norm of a Slater determinant depends upon the determinant of the corresponding orbital overlap matrix.

To summarize, the approximations we employ are equivalent to 
(1) replacing the HF Slater determinants needed to define the charge-localized
states $\Phi_{1/2}$ with Slater determinants made up of subsystem KS molecular orbitals, and (2) replacing the HF expression of $E\left[ \rho^{(12)}(\br) \right]$ by an approximate density functional.

Approximation (1) is often invoked in DFT calculations of electronic couplings \cite{kadu2012,vanv2010a,wu2006,migliore2011a,pava2011b}, while (2) was used before by Kaduk {\em et al.} \cite{kadu2012}, and it is introduced here for the first time in the context of subsystem DFT. Contrary to the density, the scaled transition density introduced in Eq.\ (\ref{supratrans}) might feature regions of space where it carries a negative sign. 
%
%
{Generally, exchange-correlation and kinetic-energy functionals
may feature arbitrary fractional powers of the density that
would render the exchange-correlation energy complex. By
working with a scaled transition density that is quasi-positive
almost everywhere in space (see below), the simplest pragmatic workaround
is to set the scaled transition density to zero everywhere
in space where it actually has negative values. While this may
seem to be a somewhat drastic approximation, we note that
in practice only a small part of the scaled transition density needs to
be changed, and the very good results observed in our test cases
below empirically justify this procedure (see Section \ref{sect_Val}).}
%
%
In all the cases considered here, the neglect of the negative parts of the scaled transition density affected its total integral by less than a tenth of a percentile point. 

%
%
{The quasi positivity of the scaled transition density can be explained with two arguments. In a frozen core approximation, if the only orbital product term left in the scaled transition density is negative almost everywhere in space, then that product will give rise to a negative element of the inverse of the transition overlap matrix (being just the inverse of that orbital overlap in this approximation). Thus, the scaled transition density will be positive almost everywhere, as the above mentioned orbital product is multiplied by the inverse overlap in the equation for the scaled transition density. The second argument uses the fact that the scaled transition density integrates to $N$, it can be decomposed into two contributions, one corresponding to the $N-1$ electron system undergoing little changes in the transition, and the other corresponding to the single transferring electron. Only the latter component has possibly negative contributions to the scaled transition density, and is unlikely to overpower the much larger $N-1$ component. 

Generally, in those cases where the $S_{12}$ overlap is zero because $x$ orbital overlaps ($x$ rows/columns of the transition overlap matrix) are identically zero, we employ the Penrose inversion of the transition overlap matrix and $\rho^{(12)}$ is indeed a $N-x$ ``positive'' electron system ``plus'' an $x$ electron system that overall integrates to zero (and thus half positive and half negative). In this case the scaled transition density integrates to $N-x$ and because $x\ll N$, the negative parts of the $x$ electron system will unlikely overpower the positive $N-x$ electron system. In any even, this orthogonal case will yield a zero $H_{12}$ matrix element as calculated by Eq.\ (\ref{newcoupling}). Thus, even in those cases where the scaled transition density has large negative parts (which we never encountered in our calculations), its contribution to the Hamiltonian matrix element will be negligible due to either a negligible or a zero overlap element.}
%
%

%
%
{We should warn the reader that the sign of the scaled transition density is independent of wave function phase factors. In fact, wave function phase factors do not affect the scaled transition density as the overlap element between the $N$-electron wave functions (Slater determinants in the context of our work) will carry with it the same phase factor as the conventional transition density. Thus, following the definition of the scaled transition density in Eq.\ (\ref{supratrans}), dividing the conventional transition density by the overlap has the effect of removing the phase factor altogether.}
%
%

For sake of brevity, hereafter we will drop ``scaled'' when referring to the above scaled transition density. In calculating the transition density from FDE subsystem orbitals coming from the KSCED equation in Eq.\ (\ref{KSCED}) by applying Eq.\ (\ref{supratrans}), one must consider all of its subsystem contributions as well as its overlap-mediated inter-subsystem couplings. The transition density for a system partitioned in $N_S$ subsystems becomes
\begin{equation}\label{newtrans}
\rho^{(12)}(\br)=\sum_{I,J}^{N_S} \sum_{k\in I}\sum_{l \in J} \phi_{(k)_I}^{(1)} (\br) \left(\mathbf S^{(12)} \right)^{-1}_{kl} \phi^{(2)}_{(l)_J} (\br).
\end{equation}
It is interesting to notice that the above transition density 
``joins'' orbital transitions of one subsystem with the ones of another subsystem through
the terms in the sum over the subsystem's labels where $I \neq J$. The coupling
of these inter-subsystem transitions are weigthed by the matrix elements of the inverse
of the transition orbital overlap matrix, $\left(\mathbf S^{(12)} \right)^{-1}$.
The diagonal density, i.e.\ the density of the broken-symmetry charge-localized states, can be formulated similarly applying Eq.\ (\ref{supradens}), namely,
\begin{equation}\label{newtransd}
\rho^{(i)}(\br)=\sum_{I,J}^{N_S} \sum_{k\in I}\sum_{l \in J} \phi_{(k)_I}^{(i)} (\br) \left(\mathbf S^{(ii)} \right)^{-1}_{kl} \phi^{(i)}_{(l)_J} (\br).
\end{equation}
As pointed out previously~\cite{pava2011b} this density is not exactly the sum of subsystem densities obtained with the FDE calculation from Eq.\ (\ref{eq:int:1}), as the intersubsystem overlap elements do not appear in the FDE formalism.
Such a density mismatch constitutes a second-order effect~\cite{pava2011b} and is not a concern here as we are going to apply the method only to  non-covalently bound subsystems, that is, super-systems with small inter-subsystems orbital overlap. As we will see in the next section, this density issue is completely by-passed by an FDE-compatible theory we call ``disjoint transition density formulation''.

\subsection{Subsystem-Joint and Subsystem-Disjoint Transition Density Formulations}
\label{sect_cu}
The transition density in Eq.\ (\ref{newtrans}) can be approximated by 
a density composed only of intra-subsystem orbital transitions. 
By recognizing that the off-diagonal blocks $S^{(12)}_{I,II}$ in Eq.\ (\ref{overdef2})
are always very small in magnitude when the subsystems are not covalently bound and if 
there is no electron transfer between subsystems 
the disjoint transition density can be written as
\begin{align}\label{newtransu} 
\tilde{\rho}^{(12)}_{DTD}(\br)&=\sum_{I} \rho^{(12)}_I (\br), \\
\label{newtransub}
\rho^{(12)}_I (\br) &=\sum_{k,l \in I} \phi_{(k)_I}^{(1)} (\br) \left(\mathbf S^{(12)}_I \right)^{-1}_{kl} \phi^{(2)}_{(l)_I} (\br),
\end{align}
where the tilde has been included to distinguish the above definition of disjoint transition density with the one we use for charge-transfer transition, see below.
The off-diagonal Hamiltonian matrix element ($H_{12}$) 
can therefore be calculated with either the joint transition density (JTD) in Eq.\ (\ref{newtrans}), giving rise to the JTD Hamiltonian couplings; or with the disjoint transition density (DTD) of Eq.\ (\ref{newtransu}), giving rise to the DTD Hamiltonian couplings.

The above definition of disjoint transition density can be cast in terms of a pure subsystem DFT formulation. Because in the FDE formalism both the ground and excited state densities can be written as a sum of subsystem densities, as in Eq.\ (\ref{eq:int:1}), namely,
\begin{align}
\rho_{\rm DTD}^{(1)}(\br) &= \sum_I \rho_I^{(1)}(\br),~\mathrm{and} \\
\rho_{\rm DTD}^{(2)}(\br) &= \sum_I \rho_I^{(2)}(\br).
\end{align}
Given that the subsystem densities are derived from Slater determinants of subsystem orbitals, then it is enough to invoke approximation (1) (i.e.\ replace the HF Slater determinants needed to define the charge-localized states $\Phi_{1/2}$ with Slater determinants made up of subsystem KS molecular orbitals) the transition density relating states 1 and 2 is of the DTD type [as in Eq.\ (\ref{newtransu})]. However, if the electronic transition involves electron transfer between subsystem $K$ and $L$, then the transition density cannot be completely disjoint, and it can be approximated by (note, compared to Eq.\ (\ref{newtransu}) the tilde is removed)
\begin{align}\label{newtransu2} 
\rho^{(12)}_{\rm DTD}(\br)&=\rho^{(12)}_{KL} (\br) + \sum_{I=1, I\ne K,L}^{N_F} \rho^{(12)}_I (\br)\\
\label{newtransu2b} 
\rho^{(12)}_{KL} (\br)&=\sum_{k,l \in K, L} \phi_{(k)_{K,L}}^{(1)} (\br) \left(\mathbf S^{(12)}_{K,L} \right)^{-1}_{kl} \phi^{(2)}_{(l)_{K,L}} (\br),
\end{align}
where $\mathbf S^{(12)}_I $ are the subsystem transition orbital overlap matrices and $\mathbf S^{(12)}_{KL}$ 
 is the full transition orbital overlap matrix of the combined $K$ and $L$ subsystems and $\rho^{(12)}_I (\br)$ is the same as in Eq.\ (\ref{newtransub}).

The diagonal elements of the Hamiltonian over the broken-symmetry states in the DTD calculations become
\begin{align}\label{uncdiag}
\nonumber
H_{ii}&=\langle \Phi_i | \hat H | \Phi_i \rangle = E\left[ \rho_{\rm DTD}^{(i)}(\br) \right] \\
&= \sum_I E\left[ \rho_I^{(i)}(\br) \right] + T_s^{\mathrm{nadd}} \left[ \rho_I^{(i)}, \rho_{II}^{(i)}, \ldots \right] +
E_{\mathrm xc}^{\mathrm{nadd}} \left[ \rho_I^{(i)}, \rho_{II}^{(i)}, \ldots \right].
\end{align}
For the above expressions, we use the FDE total energy partitioning in subsystem energies and non-additive functionals~\cite{gotz2009}. Similarly, the off-diagonal elements become
\begin{align}\label{unccoupling}
\nonumber
H_{12}&=\langle \Phi_1 | \hat H | \Phi_2 \rangle= E\left[ \rho_{\rm DTD}^{(12)}(\br) \right] S_{12} \\
&=S_{12} \left( \sum_I E\left[ \rho_I^{(12)}(\br) \right] + T_s^{\mathrm{nadd}} \left[ \rho_I^{(12)}, \rho_{II}^{(12)}, \ldots \right] +
E_{\mathrm xc}^{\mathrm{nadd}} \left[ \rho_I^{(12)}, \rho_{II}^{(12)}, \ldots \right] \right),
\end{align}
where the sums run over the number of disjoint subsystems in the calculation and the non-additive functionals have the same meaning as in the regular FDE formalism~\cite{solo2012,jaco2008b,weso1993,weso2006}. Note that for charge-transfer transitions, the above sum over the disjoint subsystems is carried out in a similar way to the one in Eq.\ (\ref{newtransu2}), i.e.\ the subsystems undergoing variation of the number of electrons [$K$ and $L$ in Eq.\ (\ref{newtransu2b})] are grouped together in one supersystem that maintains an overall constant number of electrons during the transfer of charge.

Let us remark that the computational cost needed to calculate the matrix elements in the DTD formalism is linear-scaling with the number of subsystems, while the JTD formalism is not -- it scales as $N^3$ with $N$ being the number of electrons in the supersystem even though with a much smaller scaling coefficient than regular KS calculations (as in the JTD formalism the KS equations do not need to be self-consistently solved for the supermolecular system. Instead the transition orbital overlap matrix is inverted only once).

\section{Validation of the methodology}
\label{sect_Val}

After implementation of the above described JTD and DTD algorithms in \adf, in what follows, we aim at providing a thorough benchmark of the method. We compare our calculated migration CT excitation energies against two dimer systems: a water dimer radical cation, (H$_2$O)$_2^+$, and an ethylene dimer radical cation, (C$_2$H$_4$)$_2^+$, at varying inter-monomer separations for which we carried out several different high-level wave function method calculations of the excitation energies. We also calculate hole transfer excitation energies for several DNA nucleobase dimers, and compare them to CASPT2 calculations. The idea behind this benchmark is to seek a validation of the ability of our FDE method (note JTD and DTD are equivalent for only two subsystems) to reproduce quantitatively (with at least chemical accuracy, 1 kcal/mol or 0.04 eV deviation) the CT excitation energies calculated with the wave function methods considered. 
Unless otherwise noted, all DFT calculations have been carried out with the PW91 density functional and a TZP (triple-zeta with polarization) basis set with \adf, all EOM-CCSD(T) calculations have been carried out with the {\sc NWChem 5.2} software \cite{nwchem_program}, and all Fock--Space CCSD calculations have been carried out with the {\sc DIRAC 11} software \cite{dirac_program}.
\subsection{Water Dimer}
\label{sect_Wd}
\begin{figure}
\caption{Migration charge-transfer excitations (in eV) for water dimer radical cation in the $C_{s}$ symmetric configuration
               with an intermonomer displacement from the equilibrium distance of $0<R<15$ \AA.}
\label{plot_Wd}
\begin{center}
\includegraphics[angle=-90,width=17cm]{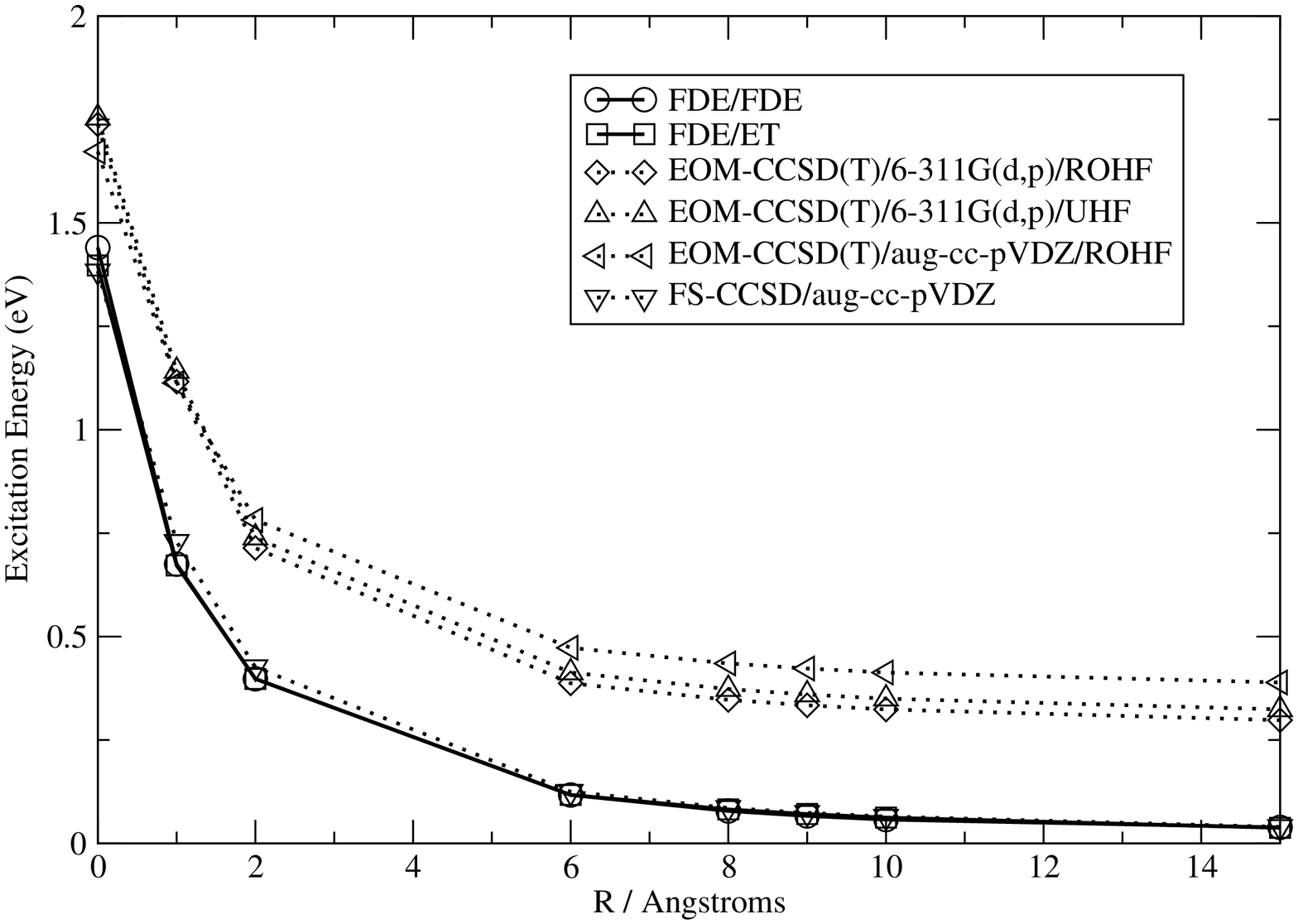}
\end{center}
\end{figure}

The nuclear positions of a neutral water dimer were first optimized with a DFT method employing the BP86 functional and a TZP basis set (triple zeta with polarization). Figure S1 \cite{epaps} depicts the obtained structure.
The subsequent calculations were carried out on the same system by translating one water molecule away from the other 
along the hydrogen-bond axis. The results are summarized in Figure \ref{plot_Wd}.
We calculated the excitation energy in two ways. In the first one (FDE/FDE label in the figure) the diabatic energies [or the diagonal elements of the diabatic Hamiltonian in Eq.\ (\ref{Hdia})] were computed using Eq.\ (\ref{uncdiag}) with the FDE density calculated as in Eq.\ (\ref{eq:int:1}). In the second calculation (FDE/ET) the diabatic energies were calculated by integrating the energy functional as in Eq.\ (\ref{newener}) with the density computed from the FDE subsystem orbitals as in Eq.\ (\ref{supradens}).  

We compare our values with excitation energies from four coupled cluster calculations. Three of these calculations were carried out with the EOM-CC method (with different basis sets and reference determinant). In the fourth calculation the excitation energy is obtained by subtracting from each other the ionization potentials obtained with Fock-Space Coupled-Cluster~\cite{viss2001} [FS-CCSD] calculated from two different guesses of the initial ionized state: one being the HOMO, almost entirely localized onto one monomer, and the other one the HOMO-1, almost entirely localized onto the other monomer.

The comparison of the CC methods with our values is very good for the FS-CCSD values while the EOM-CCSD(T) values are consistently larger in magnitude. This behavior of EOM-CC methods has been noted before for MCT excitations~\cite{glae2010} and, as mentioned in the introdution, it is due to the fact that the EOM-CC excited-state ansatz cannot be written in the regular exponential form, thus making EOM-CC non-size extensive~\cite{musi2011}. One can rationalize this with an argument based on orbital relaxation. In the excited state, the charge has migrated to another location and the orbital of the donor must relax back to the neutral state. Similarly, the orbitals of the acceptor must relax to the charged state. In both cases single excitations are the main contribution, however they must be computed on top of the single excitation needed to describe the CT itself. Therefore one needs at least double excitations to describe these relaxation effects~\cite{subo2011}. Adding perturbatively the triple excitations usually further improves this situation allowing the EOM-CCSD(T) to reach typical deviations of 0.1--0.3 eV on excitation energies~\cite{glae2010}.

We therefore consider the EOM-CCSD(T) energy overestimation a good sign that our method yields accurate excitation energies. Convincing evidence is provided by the FS-CCSD energies, as they compare with our values to within chemical accuracy at every inter-monomer separation. An issue of orthogonality arises, as the ionization potentials calculated by the FS-CCSD method are deduced from cationic wave function which are not strictly orthogonal to each other. This however, plays a very minor role here, as the overlap and the electronic coupling between the two diabatic states is almost zero as proved by our calculations of $S_{12}$ in Table S1 and S2 in the supplementary materials \cite{epaps}.

For sake of completeness, in the supplementary materials~\cite{epaps} we report the numerical values used to obtain the plot in Figure \ref{plot_Wd} as well as additional FDE calculations carried out with the BLYP functional showing a similar behavior to the PW91 calculations.
\subsection{Ethylene Dimer}
\label{sect_Ethd}
\begin{figure}
\caption{Migration charge-transfer excitations (in eV) for ethylene dimer radical cation in the $D_{2h}$  symmetric configuration
               with an intermonomer displacement from the equilibrium distance of $6<R<15$ \AA. Full $3<R<15$ \AA\ 
               range available in Table S3 \cite{epaps}.}
\label{plot_Ethd}
\begin{center}
\includegraphics[angle=-90,width=17cm]{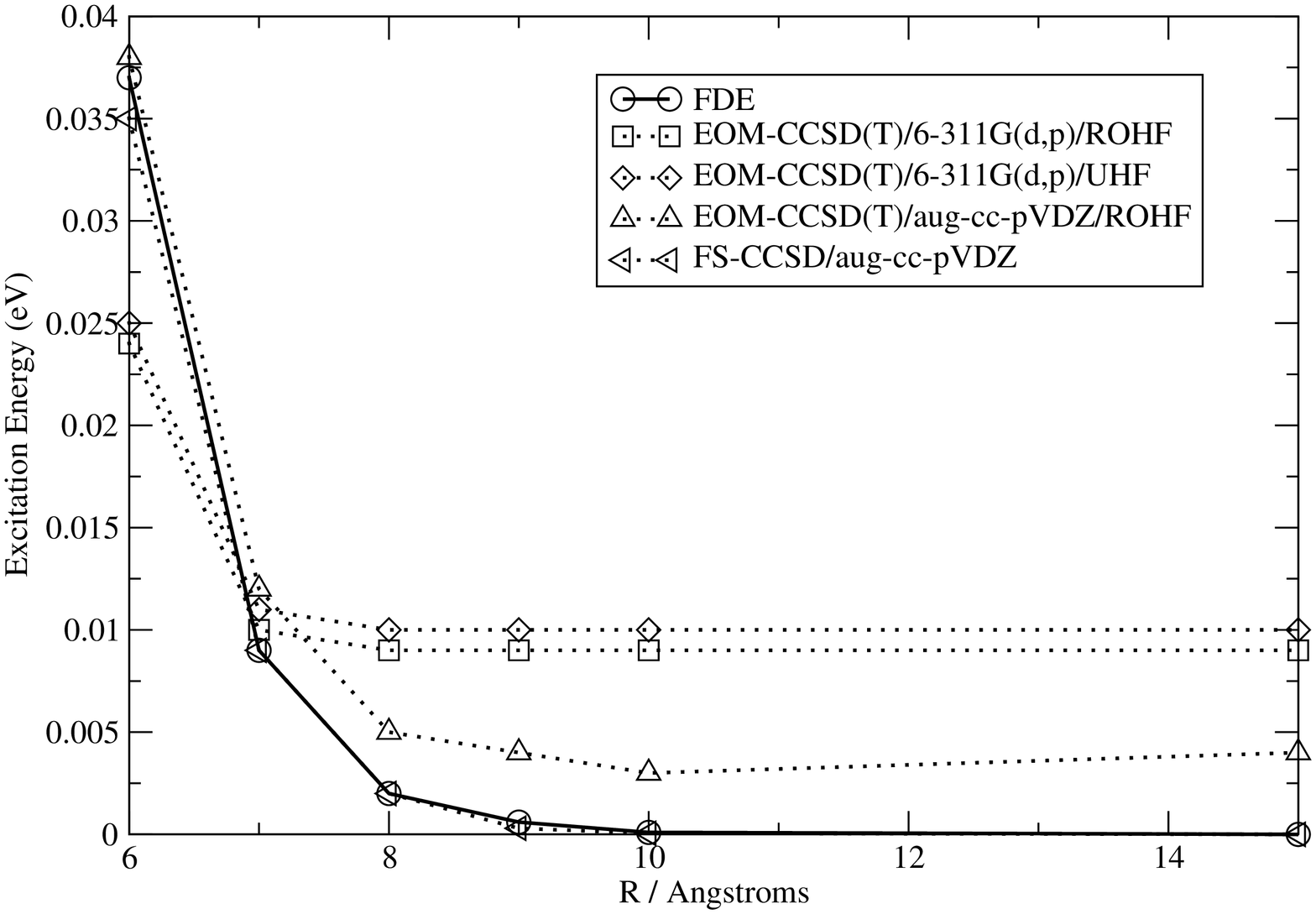}
\end{center}
\end{figure}

The nuclear positions of a neutral ethylene molecule were first optimized with a DFT method employing the BP86 functional and a TZP basis set (triple zeta with polarization). Then a copy of the same molecule is pasted so as to obtain a $\pi$-stacked ethylene dimer with intermolecular separation of $R$. Figure S2 \cite{epaps} depicts the obtained structure. The calculated excitation energies are plotted in Figure \ref{plot_Ethd}.
Due to the symmetric arrangment of the ethylene dimer, this system behaves in a very different way from the water dimer considered above. Here, the orbital relaxation issues faced before are less important as the hole is completely delocalized between the two monomers -- the inner orbitals are not dramatically affected by the excitation. The hole sits in orbitals with different symmetry in the ground and in the CT excited state. The ground state has a symmetric HOMO having $a_g$ symmetry, while the first excited state has the HOMO of $b_{2u}$ symmetry. Because of these reasons, we expect the EOM-CC calculations to be of much higher accuracy than for the water dimer case. And in fact this is what we notice. Our calculations are in very good agreement with both the EOM-CCSD(T) and the FS-CCSD calculations. Our values slightly underestimate  the CC calculation at shorter separations, even if only of 0.1 eV. At larger separations (R$>6$), due to the lowering magnitude of the excitation energy, and to a much reduced extent compared to the water dimer case, the overestimation of the excitation energy by EOM-CCSD(T) becomes noticeable again. Instead, FS-CCSD and our calculations are still in very good agreement. The EOM-CC calculations carried out with the 6-311G(d,p) basis set show a slight off-trend behavior. We attribute this to the absence of augmented functions in the basis set.

Once again, for sake of completeness, in the supplementary materials~\cite{epaps} we report the numerical values used to obtain the plot in Figure \ref{plot_Ethd} as well as additional FDE calculations carried out with the BLYP functional showing a similar behavior to the PW91 calculations.

\subsection{Hole transfer in DNA $\pi$--stacked nucleobases}
In a recent paper \cite{pava2011b} we have applied FDE to constrain charges on DNA nucleobases and subsequently we calculated electronic couplings and CT excitation energies. In the previous work we employed the formalism of Migliore {\em et al.} \cite{migliore2011a,migliore2009b,migliore2009a} for the calculation of the coupling which retained all the deficiencies of the approximate density functionals, particularly self interation. The self interaction error caused a gross overestimation of the coupling and excitation energies which was ameliorated by mixing in non-local Hartree-Fock exchange in the functional. 

Here we recalculated the same systems, i.e.\ Guanine dimer, \GG , and the Guanine-Thymine complex, \GT . In addition, here we report calculations on the Adenine dimer, \Ad , Thymine dimer, \TT , Thymine-Guanine, \TG , Adenine-Guanine, \AG , and the Guanine-Adenine, \GA .
The results are summarized in Table \ref{tab_dna}.
\begin{table}
\caption{Hole transfer excitations and couplings in $\pi$--stacked DNA nucleobase radical cation complexes. All energy values in eV. 
The -- stands for ``not available''.}
\label{tab_dna}
\begin{center}
\begin{tabular}{ccccccc}
\toprule
 Composition & $S_{12}$  & $V_{12}$ & $\Delta E_{\rm ex}^a$ &  $\Delta E_{\rm ex}^b$    &   $\Delta E_{\rm ex}^c$ & $\Delta E_{\rm ex}^d$ \\
\hline
\Ad & 0.004 & 0.092 & 0.198 & --    & --    & 0.097 \\
\AG & 0.002 & 0.177 & 0.421 & --    & --    & 0.340 \\
\GA & 0.012 & 0.058 & 0.530 & --    & --    & 0.560 \\
\GG & 0.009 & 0.051 & 0.405 & 1.653 & 0.418 & 0.392 \\
\GT & 0.020 & 0.104 & 1.082 & 1.944 & 1.159 & 1.175 \\
\TG & 0.009 & 0.056 & 0.657 & --    & --    & 0.797 \\
\TT & 0.018 & 0.099 & 0.208 & --    & --    & -- \\
\bottomrule
\end{tabular}
\end{center}
\begin{flushleft}
$a$ This work. \\
$b$ PW91/TZP from Ref.\ \cite{pava2011b} \\ 
$c$ BHandH/TZP from Ref.\ \cite{pava2011b} \\
$d$ CASPT2 values form Ref.\ \cite{voityuk2006a} 
\end{flushleft}
\end{table}
Comparing the CT excitation energies calculated with the method of Migliore with the ones calculated in this work, it is clear that the self interaction error does not affect the new calculations. This can be understood by considering that the formalism of Migliore requires the KS wave function of the supermolecular system (i.e.\ composed of the donor and acceptor). The self interaction error, even though it is present in all calculations, does not affect the FDE calculations because the hole there is localized by construction and is not allowed to overdelocalize \cite{solo2012,pava2011b}. 

The excitation energy values are generally in good agreement with the CASPT2 benchmark values. For the \TT\ system no high-level benchmark calculations are available, while for the \Ad\ and \TG\ systems our excitation energies deviate by about 0.1 eV against the CASPT2 values. 
After inspecting the excitation energies calculated with the progression CASSCF(7,8), CASSCF(11,12) and CASPT2(11,12) in Ref.\ \cite{voityuk2006a} for the \Ad\ and \TG\ systems we notice that while for all the other nucleobase stacks the excitation energies calculated with the three methods are similar to each other, the cases of \Ad\ and \TG\ stand out. For example the CASSCF(7,8) excitation energies are 0.144 eV and 1.235 eV for the \Ad\ and \TG\ systems respectively. Then these values drop to 0.047 eV and 1.097 eV for the CASSCF(11,12) and then back up to 0.097 eV for the \Ad\ system but go down to 0.797 eV for \TG\ in the CASPT2 calculation. 

According to Blancafort and Voityuk \cite{voityuk2006a}, these large fluctuations in the excitation energy are to be expected especially when going from CASSCF to CASPT2. However, they also notice that for the \Ad\ system a larger active space than they employed should be considered to confirm the couplings and excitation energies for this nucleobase dimer complex. Therefore, it is difficult to compare the excitation energies for all the nucleobase stacks we compute here with the CASSCF and CASPT2 calculations of Blancafort and Voityuk. Further analysis of these calculations (i.e.\ study the effect of the size of the active space in the CASSCF/CASPT2) lie beyond the scope of this work.
\section{Pilot calculations: embedding effects in water and ethylene clusters}
\label{sect_pilot}
In this section we present calculations carried out for systems composed of more than 2 subsystems. As pilot studies we choose the same small dimer systems considered above and embed them in clusters of the same molecule. First we consider the hole transfer in ethylene dimer embedded in an ethylene matrix, then we consider a water dimer in a cluster extracted from liquid water.

\subsection{Ethylene clusters}
\label{sect_Ethc}

\begin{figure}
\caption{Depiction of the ethylene clusters used in the calculations containing: inset (a) 4, (b) 6, (c) 8, and (d) 10 ethylene molecules. The depicted clusters were obtained from the 20-molecule cluster cutting the furthest away molecules from the center of mass of the two molecules experiencing the hole transfer. Figure obtained with {\sc molden} \cite{molden}.}
\label{fig_cEth}
\begin{center}
\begin{tabular}{m{6cm}m{6cm}}
(a) & (b) \\
\includegraphics[width=5cm]{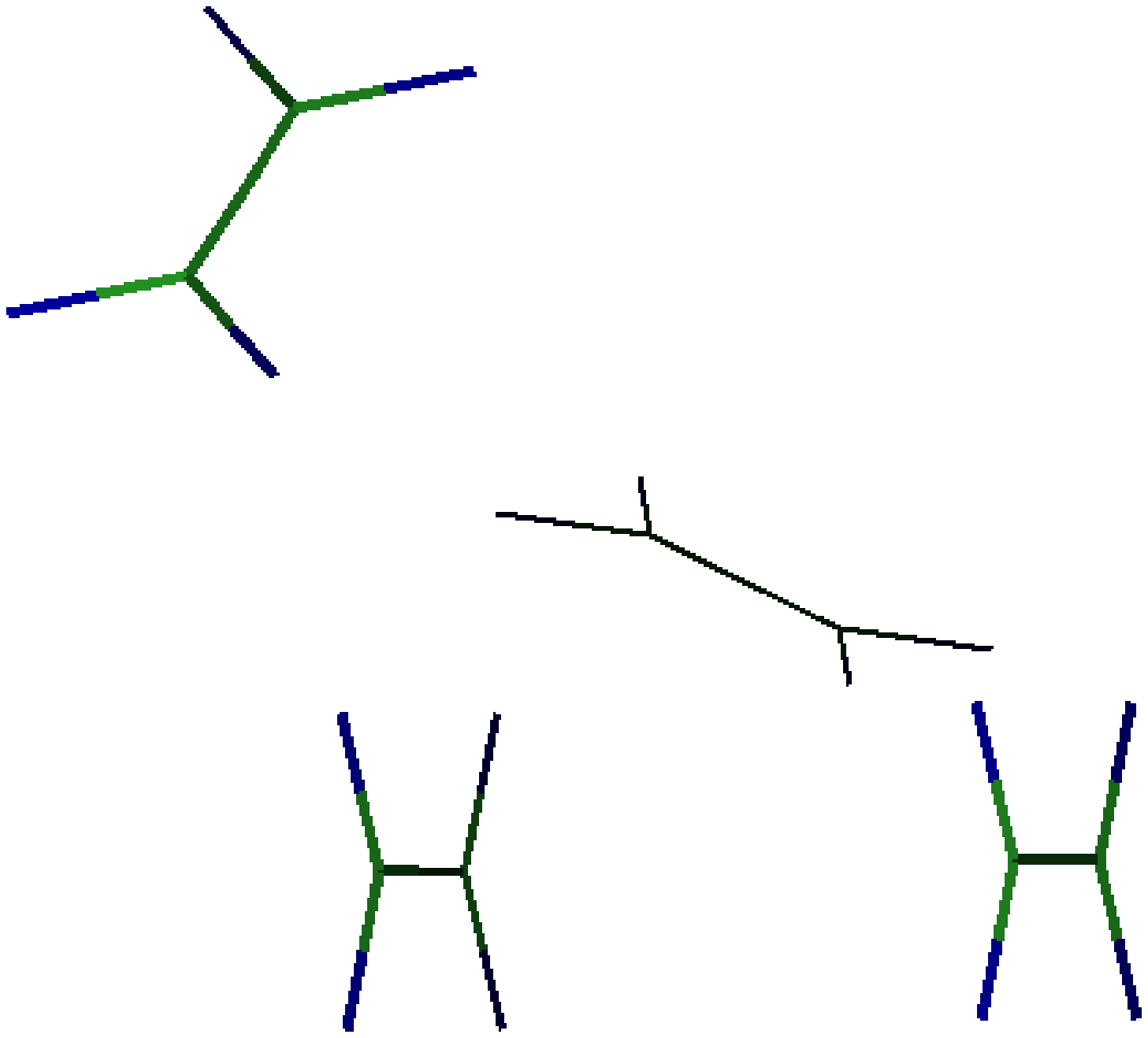} & \includegraphics[width=6cm]{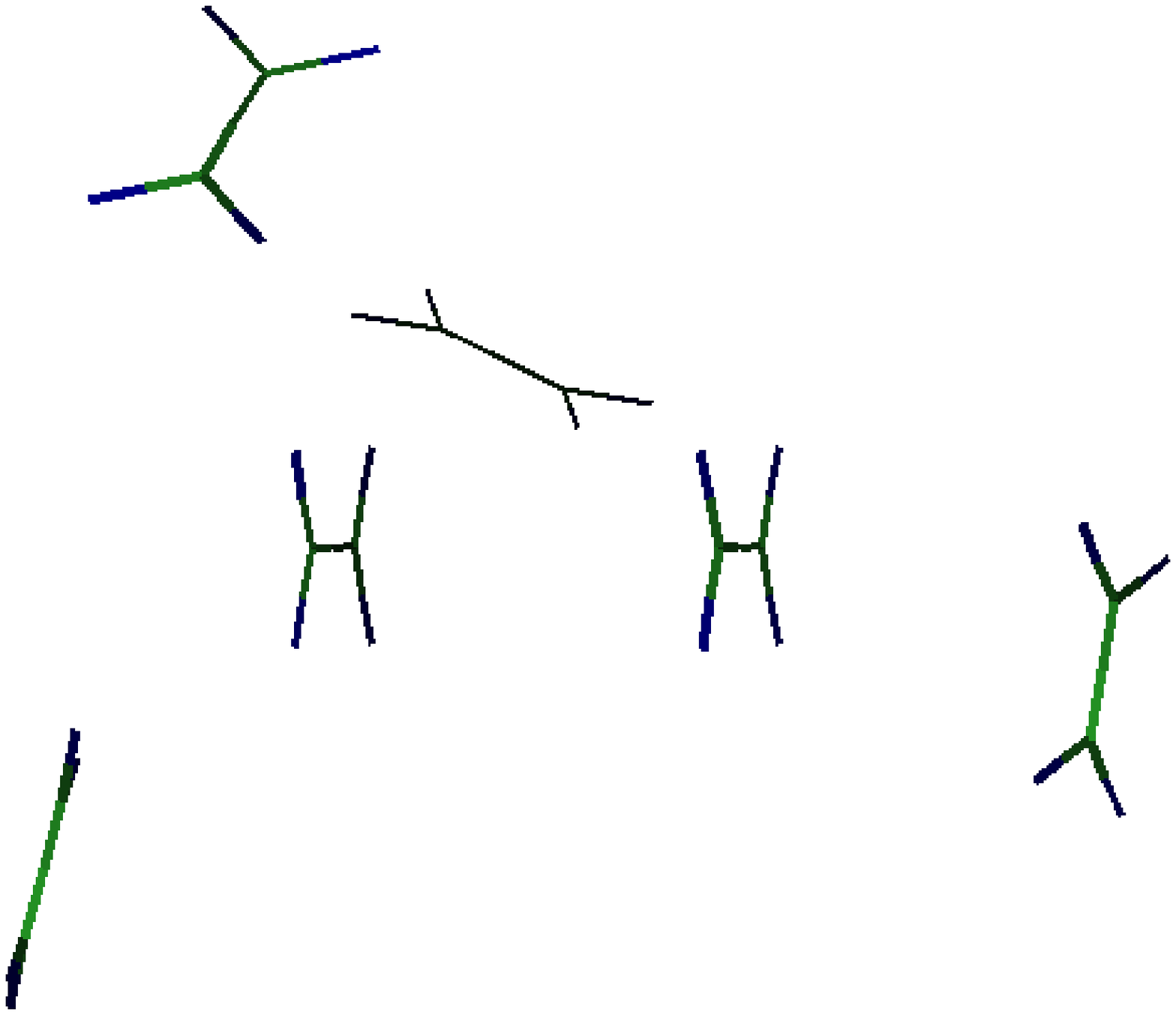} \\
(c) & (d) \\
\includegraphics[width=6cm]{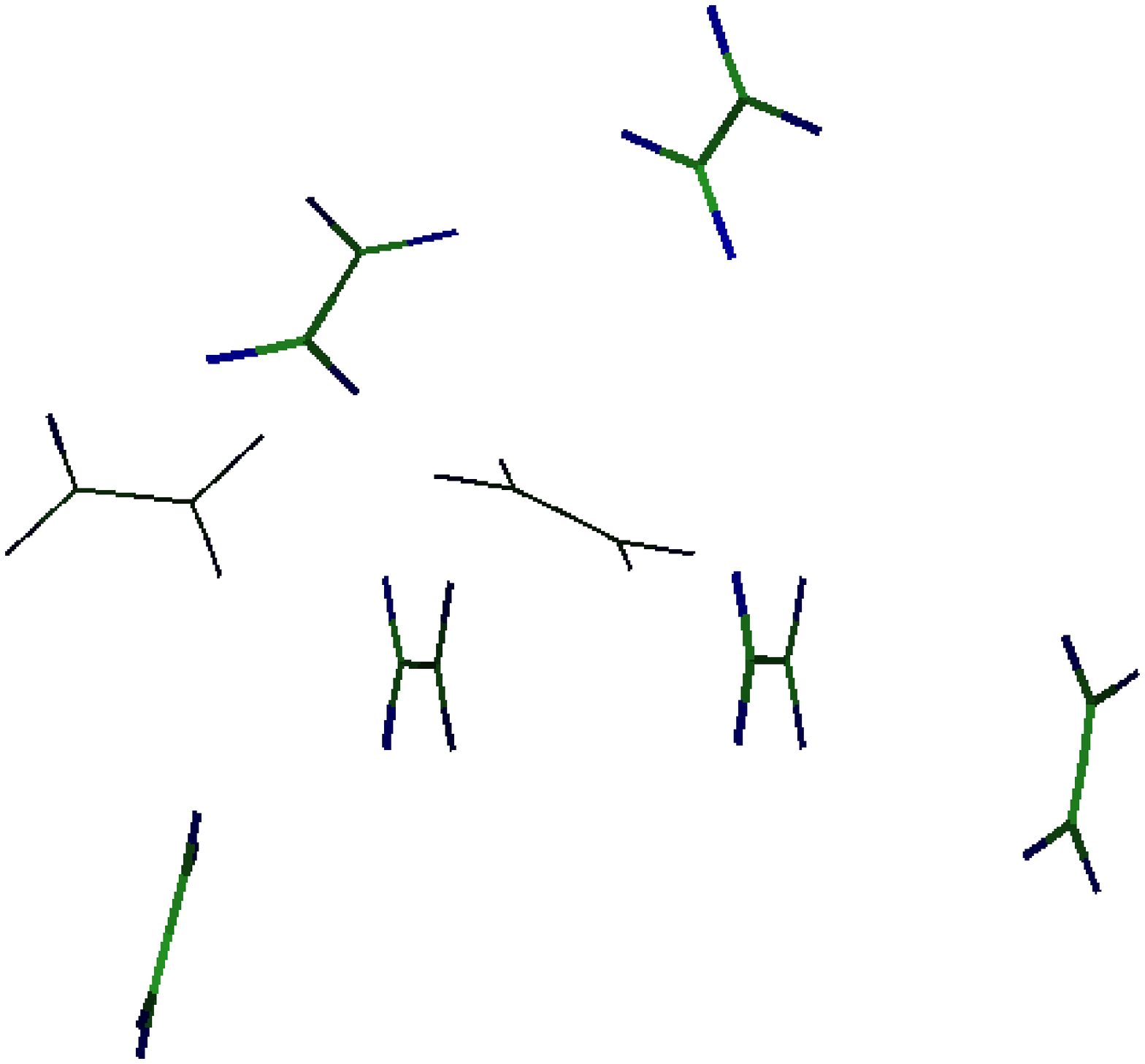} & \includegraphics[width=6cm]{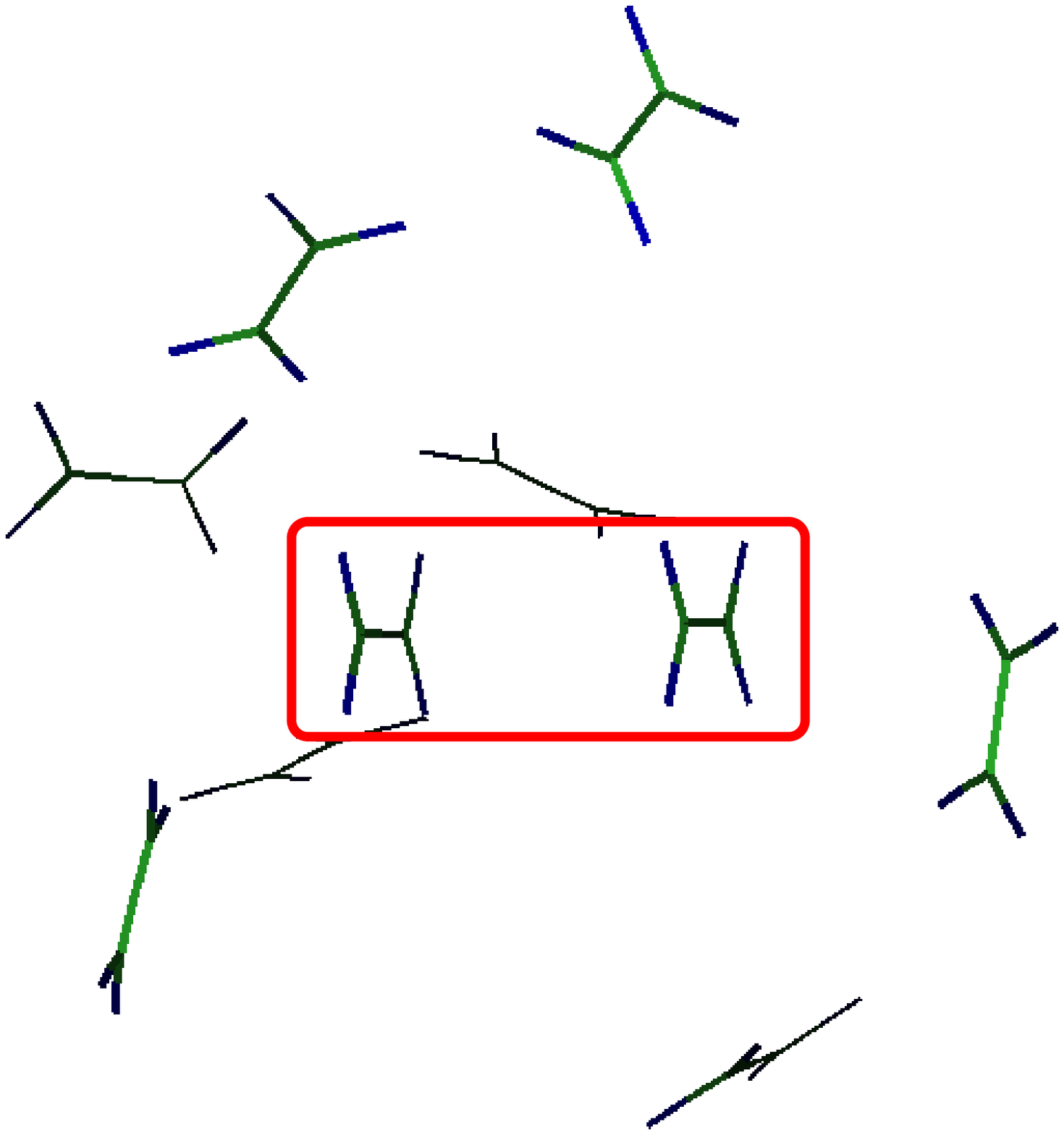} \\
\end{tabular}
\end{center}
\end{figure}

In Figure \ref{fig_cEth} we depict a selected number of ethylene clusters used in the calculation, and in Table \ref{tab_cEth} we report the values of excitation energy and electronic coupling corresponding to the hole transfer in the selected dimer. The dimer has the same geometry of the dimer considered in Section \ref{sect_Ethd} with inter-monomer separation of $R=4.0$ \AA . For the DTD calculation the non-additive exchange-correlation functional used is PW91, while PW91k~\cite{weso1996,lemb1994} was employed for the evaluation of the non-additive kinetic-energy functional.
\begin{table}
\caption{Ethylene cluster electronic couplings and charge transfer excitations. 
         The hole transfer occurs between two ethylene molecules separated by $R=4.0$ \AA .
         All values in eV. $N$ is the number of molecules in the cluster.}
\label{tab_cEth}
\begin{center}
\begin{tabular}{rcccc}
\toprule
  & \multicolumn{2}{c}{JTD}    & \multicolumn{2}{c}{DTD}     \\
$N$ & $V_{12}$ & $\Delta E_{\rm ex}$ & $V_{12}$ & $\Delta E_{\rm ex}$ \\
\hline
2   & 0.260 & 0.521 & ----- & ----- \\
4   & 0.261 & 0.539 & 0.261 & 0.540 \\
6   & 0.262 & 0.524 & 0.260 & 0.521 \\
8   & 0.262 & 0.535 & 0.261 & 0.534 \\
10  & 0.262 & 0.538 & 0.260 & 0.538 \\
20  & 0.262 & 0.535 & 0.260 & 0.534 \\
\bottomrule
\end{tabular}
\end{center}
\end{table}
The excitation energy and coupling values reported in Table \ref{tab_cEth} are essentially independent of the size of the cluster, showing that for this dimer system we should expect a minor environmental effect of the CT excitation and the hole transfer electronic coupling. This is indeed what was found for this conformation and inter-monomer orientation in a study by Lipparini and Mennucci~\cite{lipp2007} where the environment was modelled as a polarizable dielectric.

It is important to notice how the JTD and DTD calculations yield very similar excitation energies and couplings. The DTD values always lie a few meV lower than the JTD ones.  The RMS deviation of the excitation energies calculated with the JTD versus DTD method is 1.5 meV, while the average deviation is 1.2 meV. While it is difficult to pinpoint the origin of this small discrepancy, we note that the contribution to the excitation energy and electronic coupling from the non-additive kinetic energy functional ranged from only a few meVs in most cases to 23 meV in the four-member cluster system. In all cases, this contribution brought the excitation energy value closer to the JTD one.

Lipparini and Mennucci~\cite{lipp2007}, also considered dimers with different orientation by rotating one of the ethylene molecules around the carbon-carbon axis. In the supplementary materials \cite{epaps}, we report calculations carried out for the dimer systems with a rotated monomer in the presence of a minimal environment. Our calculations largely reproduce the findings of Lipparini and Mennucci. However, our FDE calculations allow for the recovering of effects of the discrete molecular environment.

\subsection{Water clusters}
\label{sect_Wc}

The clusters were generated by chosing a water dimer complex from bulk water and then carving the desired water molecules around it. The coordinates of a large cluster of bulk water was given to us by Daniel Sp{\aa}ngberg~\cite{spanpriv}. More details on the generation of the bulk water is available in the supplementary material\cite{epaps}.

Figure \ref{fig_cW} depicts a selected set of clusters considered with the highlighted water dimer chosen for the hole transfer. All the other molecules belong to the environment and polarize according to where the hole is located. However, they do not undergo variation of electron number during the hole transfer process. In Table \ref{tab_cW} we collect the values of electronic couplings and excitation energies calculated with the JTD and DTD formalisms. As noted in the ethylene dimer calculations, also here the DTD values always underestimate the JTD ones. In this case by 1--60 meV for excitation energies and 1--15 meV for the electronic couplings. The RMS deviation of the excitation energies calculated with the JTD versus DTD method is 0.06 eV, while the average deviation is 0.05 eV.

In our water cluster calculation we see a very different behaviour from the ethylene cluster. The electronic couplings are as effected by the environment as the excitation energies. This is likely the effect of water having a permanent dipole, thus adding large contributions to the electric field interacting with the hole. Contrary, in the ethylene clusters case the environmental electric field acting on the hole is of much weaker magnitude as it is due only to polarization of the ethylene molecules. What we notice in our calculations is that the more water molecules are included in the cluster, the more the electronic coupling decreases. This was also a feature of the effect of embedding by water on excitonic couplings in $\pi$-stacked chromophore diads~\cite{neug2010}. For the configuration considered, the excitation energy decreases by about 1 eV as the cluster size increases. Such large variations of the excitation energy in condensed phases with polar solvents is well known~\cite{menn2006}. Witness of this are also the much larger reorganization energies of electron transfer processes in polar solvent than in non-polar ones~\cite{Nitzan_book,guti2009,seid2009,blum2008,pava2010a}.

\begin{figure}
\caption{Depiction of the water clusters used in the calculations containing: inset (a) 2, (b) 4, (c) 6, (d) 8, and (e) 10 water molecules. The depicted clusters were obtained from the 56-molecule cluster cutting the furthest away molecules from the center of mass of the two molecules experiencing the hole transfer. Figure obtained with {\sc vmd} \cite{vmd} and {\sc molden} \cite{molden}.}
\label{fig_cW}
\begin{center}
\begin{tabular}{m{6cm}m{6cm}}
(a) & (b) \\
\includegraphics[width=5cm]{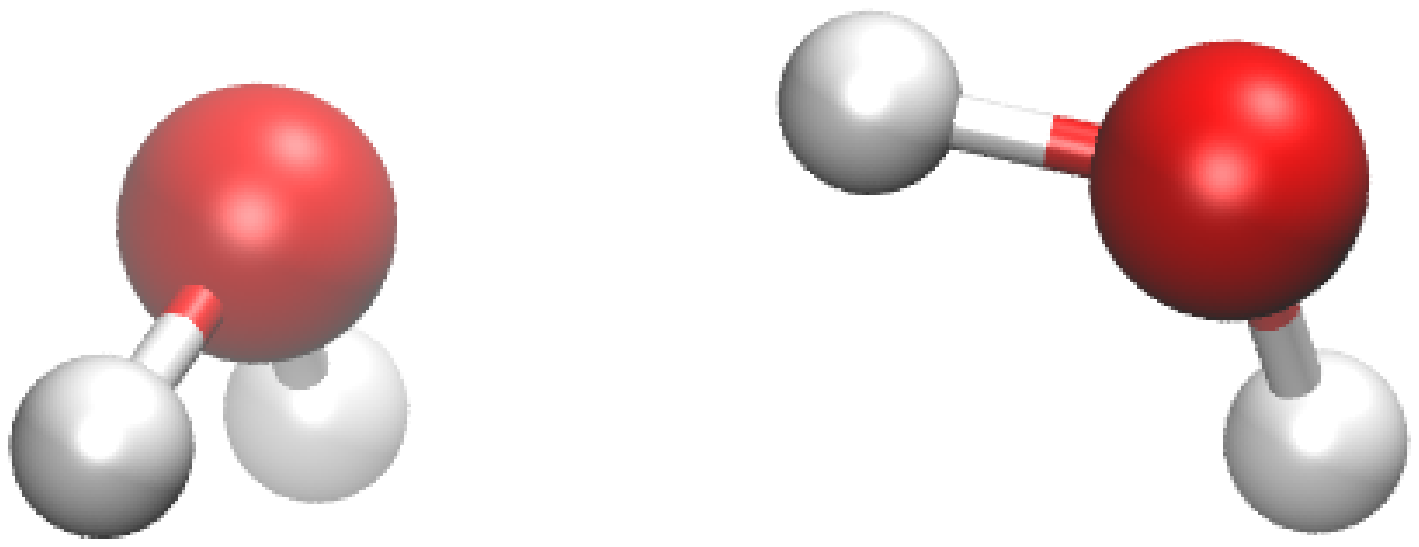} & \includegraphics[width=6cm]{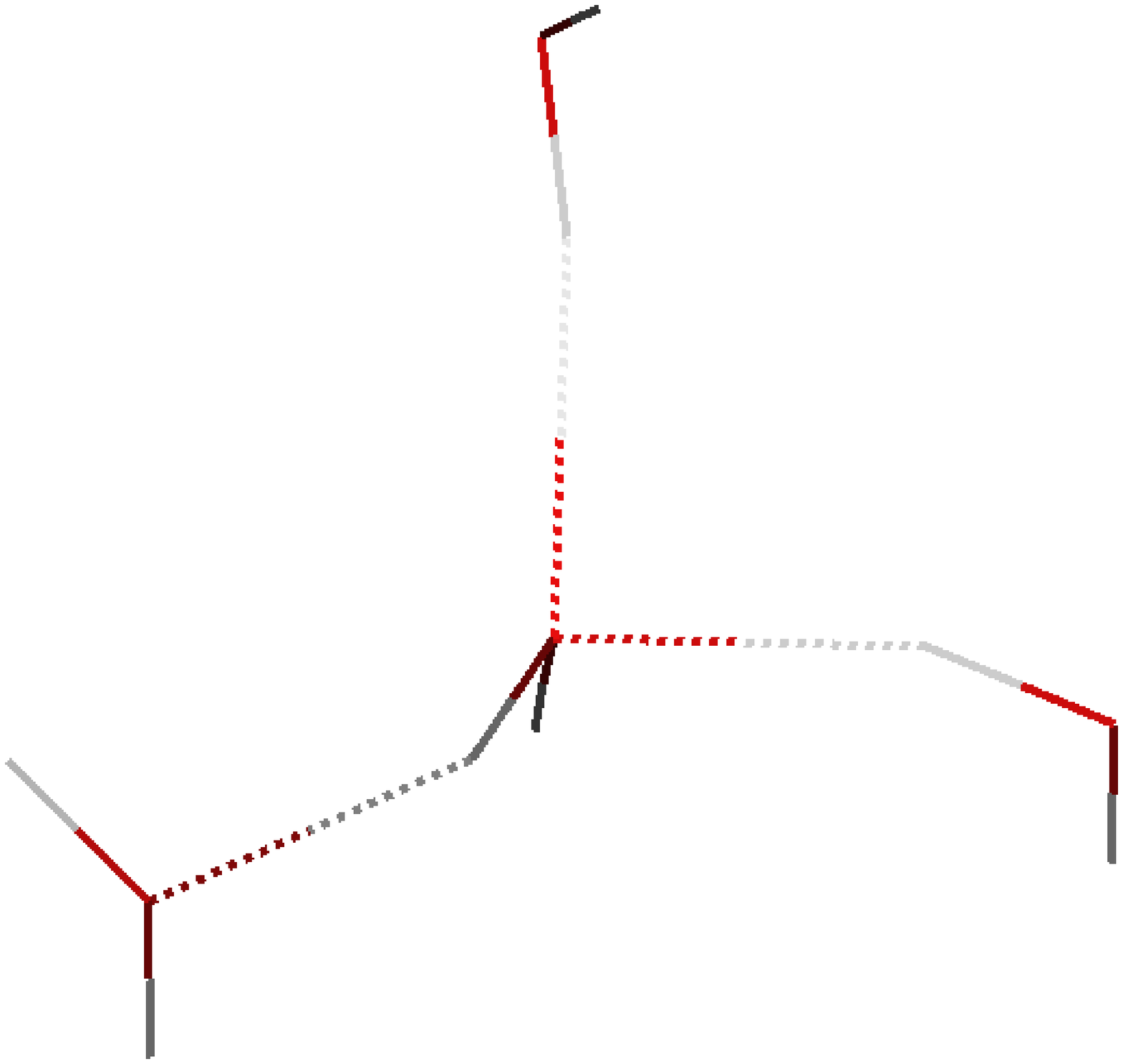} \\
(c) & (d) \\
\includegraphics[width=6cm]{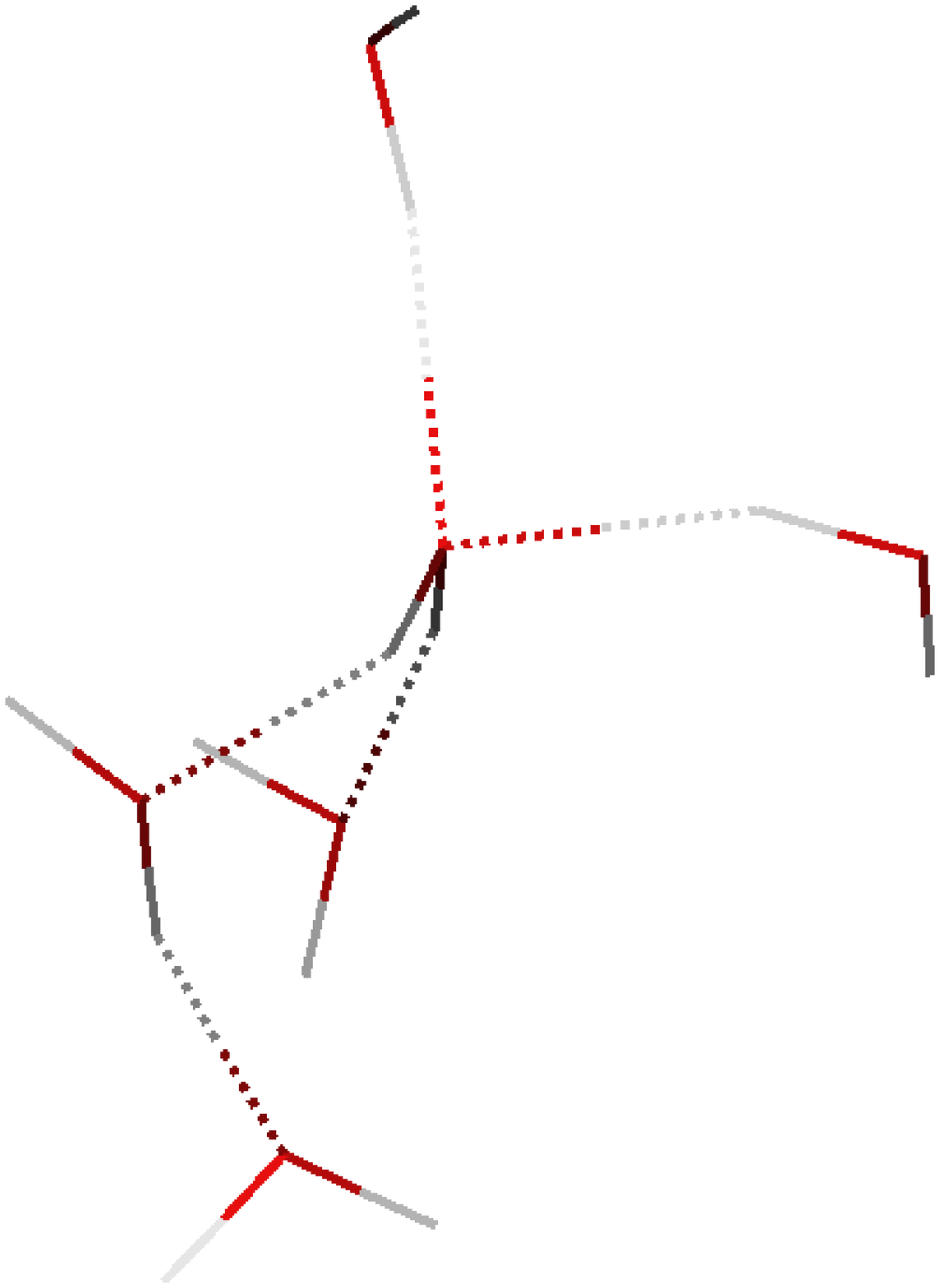} & \includegraphics[width=6cm]{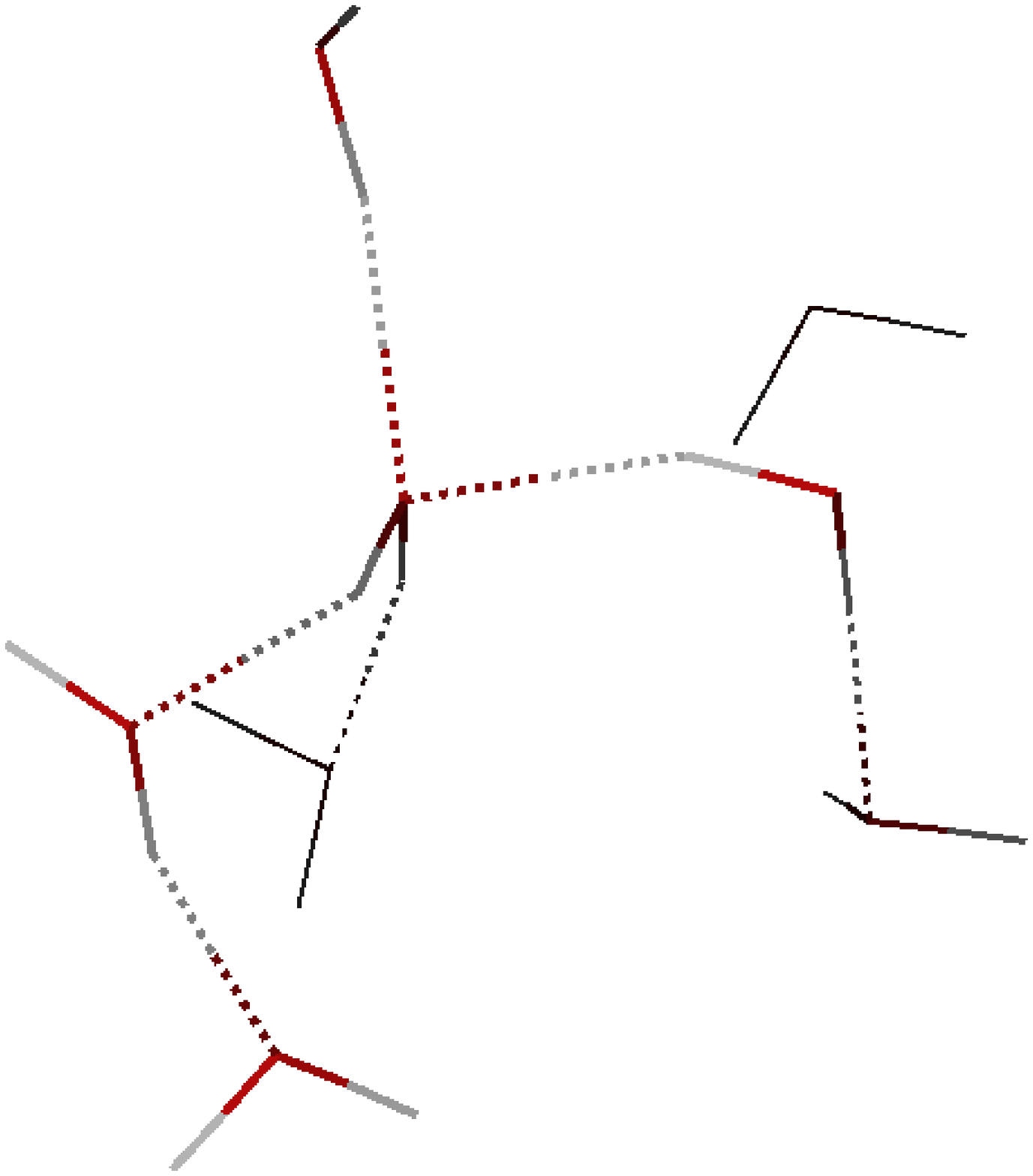} \\
\multicolumn{2}{c}{(e)} \\
\multicolumn{2}{c}{\includegraphics[width=6cm]{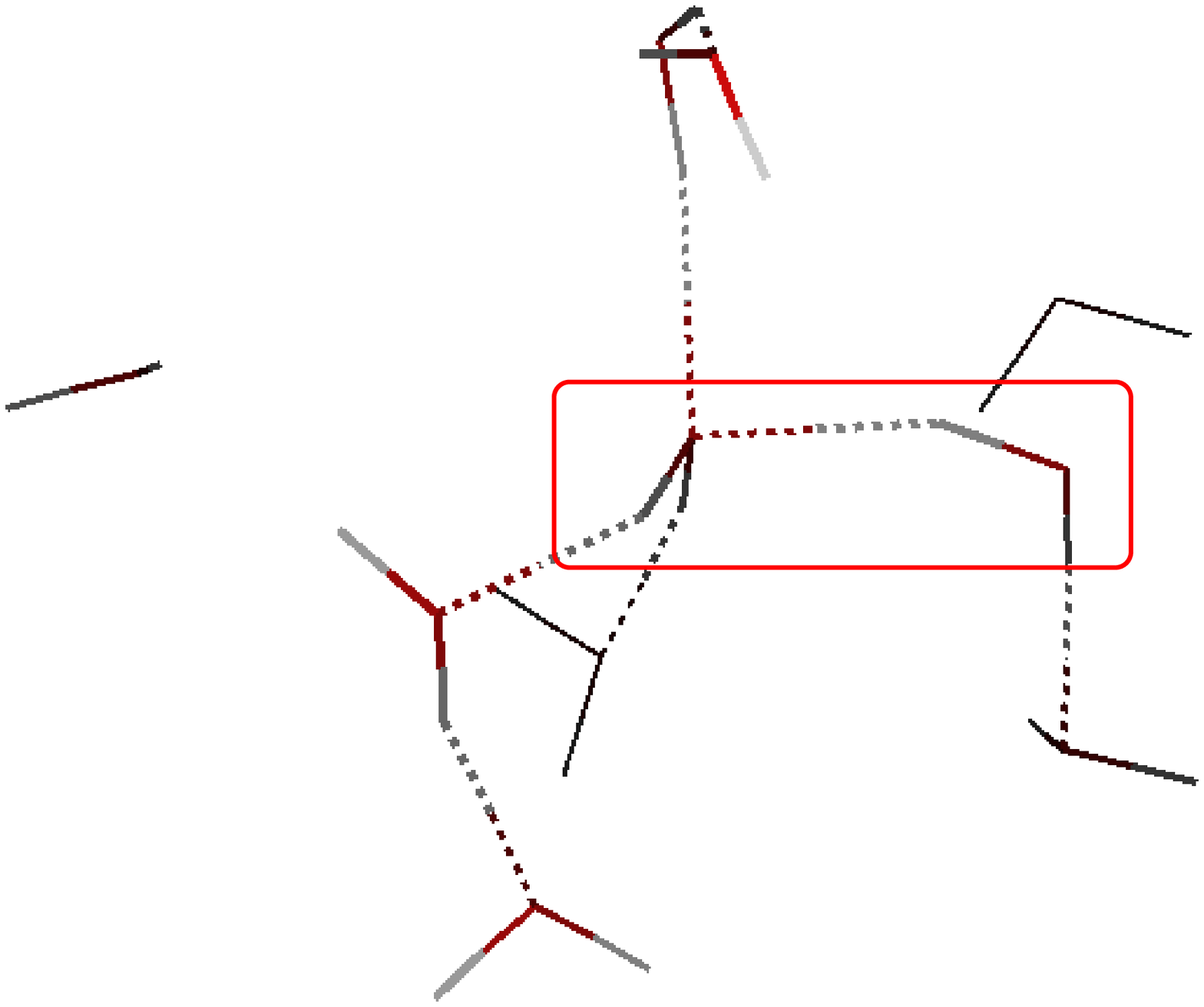} }\\
\end{tabular}
\end{center}
\end{figure}
\begin{table}
\caption{Water cluster electronic couplings and charge transfer excitations. 
         All values in eV. $N$ is the number of molecules in the cluster.}
\label{tab_cW}
\begin{center}
\begin{tabular}{rcccc}
\toprule
    & \multicolumn{2}{c}{JTD}       & \multicolumn{2}{c}{DTD}     \\
$N$ & $V_{12}$ & $\Delta E_{\rm ex}$ & $V_{12}$ & $\Delta E_{\rm ex}$ \\
\hline
2  & 0.259 & 1.431 & ----- & ----- \\
3  & 0.242 & 1.138 & 0.253 & 1.098 \\
4  & 0.250 & 1.447 & 0.249 & 1.422 \\
5  & 0.226 & 1.254 & 0.241 & 1.189 \\
6  & 0.251 & 0.951 & 0.250 & 0.913 \\
7  & 0.227 & 0.771 & 0.242 & 0.716 \\
8  & 0.257 & 1.527 & 0.252 & 1.498 \\
9  & 0.236 & 1.344 & 0.244 & 1.275 \\
10 & 0.261 & 1.416 & 0.256 & 1.383 \\
11 & 0.248 & 1.252 & 0.254 & 1.187 \\
13 & 0.249 & 1.227 & 0.255 & 1.154 \\
17 & 0.245 & 1.094 & 0.251 & 1.013 \\
21 & 0.211 & 1.002 & 0.219 & 0.929 \\
27 & 0.160 & 0.610 & 0.170 & 0.544 \\
31 & 0.177 & 0.743 & 0.186 & 0.680 \\
36 & 0.196 & 0.843 & 0.203 & 0.777 \\
41 & 0.199 & 0.733 & 0.205 & 0.700 \\
46 & 0.155 & 0.563 & 0.165 & 0.518 \\ 
51 & 0.147 & 0.607 & 0.160 & 0.558 \\ 
56 & 0.128 & 0.468 & 0.143 & 0.422 \\
\bottomrule
\end{tabular}
\end{center}
\end{table}

\section{Conclusions}
In this work, we have developed a new and linear-scaling DFT method aimed at accurately predicting charge-transfer excitation energies and diabatic couplings among charge-localized states, and validated it against high-level wave function calculations.

The success of the disjoint transition density method (or DTD, see section \ref{sect_cu}) resides in two important properties. First, it computationally scales linearly with the number of non-covalently bound molecules in the system. Secondly, the hole transfer excitation energies calculated with mainstream GGA-type functionals reproduce within chemical accuracy calculations carried out with several types of Coupled Cluster methods. For the benchmark cases considered, our method outperforms the biased EOM-CCSD(T), while it is in excellent agreement with vertical ionization potential differences calculated with Fock-Space CCSD(T) calculations. We also carried out calculations of the hole transfer excitations in seven DNA base pair combinations reproducing CASPT2(11,12) calculations within a 0.1 eV deviation or better. Pilot calculations on molecular clusters of water and ethylene have also been carried out with cluster sizes up to 56 molecules. We show how the method is able to recover solvation effects on diabatic couplings and site energies (needed in electron transfer calculations) as well as the excitation energies at the discrete molecular level and fully quantum-mechanically. This constitutes an important step forward in the development of linear scaling electronic structure methods tailored to electron transfer phenomena. 

Future improvements of the method are underway and will include the implementation of hybrid functionals in the post-SCF calculation of the full-electron Hamiltonian matrix elements and the extension of the method to both excitations involving charge separation and covalently bound subsystems.
\section{Acknowledgments}
We are indebted to Daniel Sp{\aa}ngberg for providing us with the water cluster structures. We also acknowledge Benjamin Kaduk for helpful tips for coding of the transition density, and Alexander Voityuk for helpful discussions. M.P.\ acknowledges partial support by the start-up funds provided by the Department of Chemistry and the office of the Dean of FASN, Rutgers-Newark. J.N. is supported by a VIDI grant (700.59.422) of the Netherlands Organization for Scientific Research (NWO).
%

\end{document}